\documentclass[showpacs,preprintnumbers,amsmath,amssymb,aps]{revtex4}

\usepackage{graphicx}
\usepackage{dcolumn}
\usepackage{bm}
\begin{document}

\preprint{CD2dBS-v2}

\title{Convergence dynamics of 2-dimensional isotropic and anisotropic Bak-Sneppen models}

\author{Burhan Bakar}
 \email{burhan.bakar@mail.ege.edu.tr}
\author{Ugur Tirnakli}
 \email{ugur.tirnakli@ege.edu.tr}
\affiliation{
Department of Physics, Faculty of Science, Ege University, 35100 Izmir, Turkey
}

\date{\today}

\begin{abstract}
The conventional Hamming distance measurement captures only the short-time dynamics of the displacement between the uncorrelated random configurations. The minimum difference technique introduced by Tirnakli and Lyra [Int. J. Mod. Phys. C {\textbf14}, 805 (2003)] is used to study the short-time and long-time dynamics of the two distinct  random configurations of the isotropic and anisotropic Bak-Sneppen models on a square lattice. Similar to $1$-dimensional case, the time evolution of the displacement is intermittent. The scaling behavior of the jump activity rate and waiting time distribution reveal the absence of typical spatial-temporal scales in the mechanism of  displacement jumps used to quantify the convergence dynamics.    
\end{abstract}


\maketitle
\section{\label{sec:Int}Introduction}
The term self-organized criticality (SOC) was first introduced by Bak \emph{et al.} in 1987 \cite{Bak-PRL59}. In their well-known paper, it was argued that the dynamics which gives rise to the power-law correlations seen in the non-equilibrium steady states must not involve any fine-tuning of parameters. Namely, the systems under their natural evolution are driven to a statistically stationary state where long range spatio-temporal fluctuations are seen similar to those in equilibrium critical phenomena. 

The Bak-Sneppen (BS) model, which models the biological evolution of an ecology of interacting species \cite{Bak-PRL71} is one of the simplest model that exhibits SOC. Besides its simplicity, this prototype of SOC shows a quite rich variety of properties from the critical phenomena. In the BS model, an ecosystem consisting of $L^{d}$ species on a $d$-dimensional lattice is characterized by random fitness values, each of which represents the adaptation of each species to the environment.         

In this paper, we consider only $2$-dimensional case. The dynamics of BS model on a $2$-dimensional lattice of an edge size $L$ has simple rules. The initial state of the system on a square lattice is characterized by $N=L\times L$ fitness values (random numbers) $f_{i,\,j}$, where $i=1\cdots L$ and $j=1\cdots L$, uniformly distributed between $0$ and $1$. These fitness values are assigned to each site $(i,\,j)$ of the lattice with periodic boundary conditions. The usual dynamics of the system is eventually achieved by localizing the lattice site with the minimum fitness $f_{min}$ and assigning new random numbers to that site and its first nearest neighbors. 

The model described above can be called as an \emph{isotropic} BS model since the interaction between the current minimum and its first nearest neighbors is the same in both directions. In other words, if one observes the minimum fitness value of the system at time step $t$ as $f_{min}=f_{i,\,j}$ then the fitness values $f_{i-1,\,j}$, $f_{i,\,j-1}$, $f_{i,\,j}$, $f_{i+1,\,j}$ and $f_{i,\,j+1}$ will be updated at that time step. This means that the possibility for the minimum to jump to one of its left or right nearest neighbors and to one of its up or down nearest neighbors in the next time step is simply the same, with a resulting isotropic avalanche of events. On the other hand, one can easily consider alternative updating rules. One possible alternative could be to update at each time step $f_{i-u,\,j}$, $f_{i,\,j-\ell}$, $f_{i,\,j}$, $f_{i+b,\,j}$ and $f_{i,\,j+r}$ (where $u$, $\ell$, $b$ and $r$ are arbitrary positive integers taken from the interval $[1,\,L]$). Let us consider, for example, the case of  $f_{i-2,\,j}$, $f_{i,\,j-1}$, $f_{i,\,j}$, $f_{i+1,\,j}$ and $f_{i,\,j+2}$. The system now has an inherent bias to the up and right. This means that we would expect a preferred direction for an avalanche to propagate. Therefore, this model is called as the \emph{anisotropic} BS model. In principle, several types of anisotropy can be introduced  by changing the values of $u$, $\ell$, $b$ and $r$, provided that $\ell\neq r$ and $u\neq b$ (since $\ell=r$ and $u=b$ represents the isotropic BS model). The maximal anisotropic cases are defined by $\ell=0$ ($\ell=1$), $r=1$ ($r=0$) and $b=0$ ($b=1$), $u=1$ ($u=0$) whereas other definitions are considered as intermediate anisotropies. In this work, we use the maximal anisotropic case in all simulations since the convergence is faster than the intermediate anisotropy choices \cite{Head-JPA31}. 
 
After some transient time which depends on the size of the system the isotropic and anisotropic models achieve a statistically stationary state (i.e., self-organized critical state) in which the density of fitness values is uniformly distributed on $[f_{c},\,1]$ and vanishes on $[0,\,f_{c}]$, where the critical threshold value $f_{c}$ depends on the lattice dimension. On the 2-dimensional lattice considered here, $f_{c}\simeq0.328$ \cite{Bak-PRE53} for the isotropic BS model and $f_{c}\simeq0.439$ \cite{BB-EPJ} for the anisotropic BS model. The difference between the critical threshold values of isotropic and anisotropic BS models comes from the change in the rates of the spreading out of avalanches. Once the stationary state is achieved, the temporal and spatial correlation functions are power-law in both models signifying the existence of a critical state with no characteristic length or time scales (scale invariance). These correlations can be used to determine the universality classes of such models. The distribution of the absolute distance $x$ between successive minima is a good example for the spatial correlation and defined by $P_{jump}(x)\propto x^{-\pi}$, where $\pi\simeq2.92$ for the isotropic BS model and $\pi\simeq2.57$ for the anisotropic BS model on a square lattice. The temporal correlations $P_{first}(t)$, the distribution of first return times, and $P_{all}(t)$, the distributions of all return times scale as $P_{first}(t)\propto t^{-\tau_{first}}$ and $P_{all}(t)\propto t^{-\tau_{all}}$, where $\tau_{first}\simeq1.24$ and $\tau_{all}\simeq0.71$ for $2$-dimensional isotropic BS model; $\tau_{first}\simeq1.32$ and $\tau_{all}\simeq0.85$ for $2$-dimensional anisotropic BS model. These different values of $\pi$, $\tau_{first}$ and $\tau_{all}$ suggest that the isotropic and anisotropic BS models belong to different universality classes. 

The SOC feature of the BS model is revealed in its ability to naturally evolve towards a scale invariant stationary state where the correlation length is infinite and an initial local perturbation might lead to a global effect. Therefore, the sensitivity to the initial conditions in the BS model is a crucial matter. In order to study the effects of an initial perturbation one can borrow the damage spreading technique from the dynamical systems theory, which has been used in the literature to investigate the propagation of local perturbations in $1$-dimensional \cite{Tamarit-EPJB1,Cafiero-EPJB4,Valleriani-JPA32,Cafiero-EPJB7,UT-Physica344,UT-IJMPC14,UT-Physica342} and $2$-dimensional \cite{BB-EPJ} BS models. The algorithm of this technique for a $2$-dimensional BS model considered in this work can be described as follows: (i) once the stationary state has been achieved, consider the configuration as replica $1$ denoted by $f_{i,\,j}^{1}$, (ii) produce an identical copy of  $f_{i,\,j}^{1}$ and introduce a small damage in this copy by interchanging the site with minimum fitness with a randomly chosen site (denote the new replica as $f_{i,\,j}^{2}$), (iii) let both replicas evolve in time using always the same set of random numbers. One can then define the Hamming distance between two replicas as,
\begin{equation}
\label{eq:Hamming}
D(t)=\left\langle\frac{1}{N}\sum_{i,\,j=1}^{L}| f_{i,\,j}^{1}-f_{i,\,j}^{2}|\right\rangle,
\end{equation} 
where $\langle\cdots\rangle$ stays for the configurational averages over various realizations. Measuring the evolution of the discrepancy between two initially close configurations under the same external noise shows that Eq.~(\ref{eq:Hamming}) exhibits an initial power-law divergence proportional to $t^{\alpha}$, where $\alpha\simeq0.48$ and $\alpha\simeq0.53$ for  $1$-dimensional isotropic and anisotropic BS models \cite{UT-Physica344}, and $\alpha\simeq0.83$ and $\alpha\simeq0.91$ for $2$-dimensional isotropic and anisotropic BS models \cite{BB-EPJ}, followed by a finite size dependent saturation regime. These scaling exponent values of the Hamming distance suggest that $1$- and $2$-dimensional isotropic and anisotropic BS models are weakly sensitive to the initial conditions (i.e., $\alpha>0$). 

The above mentioned Hamming distance measurement captures the short-time dynamics of the model very well. However, Tirnakli and Lyra have recently introduced a new Hamming distance measurement technique to identify both the short-time and long-time dynamics of the model \cite{UT-IJMPC14}. Their Hamming distance definition on a chain has the following form, 
\begin{equation}
\label{eq:Hamming-UT}
D_{j}(t)=\left\langle\frac{1}{L}\sum_{i=1}^{L}| f_{i}^{1}-f_{i+j}^{2}|\right\rangle,
\end{equation}
where $L$ is the system size and $j=0,\,1,\,\cdots,\,L-1$. Namely, Eq.~(\ref{eq:Hamming-UT}) says that the measure of the Hamming distance $D(t)$ between two configurations is the smallest among the $L$ possible values of $D_{j}(t)$. It should be noticed that for $j=0$ (i.e., $D_{0}(t)$) Eq.~(\ref{eq:Hamming-UT}) turns to the conventional Hamming distance measurement. This new definition of the Hamming distance has been used to study the convergence dynamics of two independent configurations of $1$-dimensional BS model \cite{UT-Physica375}. 

Our task will be to study the convergence dynamics of two uncorrelated configurations of the BS model by generalizing the recently introduced Hamming distance measurement to $2$-dimensional case. The numerical procedure and the generalization of the new Hamming distance measure to $2$-dimensional BS model are given in Section~\ref{sec:Model}. The short-time and long-time dynamics of distinct configurations of $2$-dimensional isotropic and anisotropic BS models are studied under the same external noise influence. Once the model and the numerical procedure are introduced then the statistical properties of the convergence dynamics of two uncorrelated configurations of the isotropic and anisotropic BS models constructed on a square lattice are discussed through the jump activity rate and the waiting time distributions in Section~\ref{sec:Activity}. A summary of the results in Section~\ref{sec:Conc} concludes the paper.        
\section{\label{sec:Model} Numerical procedure and time evolution of Hamming distance}
To investigate the convergence dynamics of two uncorrelated random configurations of $2$-dimensional isotropic and anisotropic BS models we implement the BS algorithm on a square lattice with periodic boundary conditions. $N=L\times L$ fitness values $f_{i,\,j}$ (random numbers taken from a uniform distribution in the interval $[0,\,1]$) are initially assigned to all sites of a square lattice of an edge size $L$. As mentioned in Section~\ref{sec:Int}, the system evolution is standard re-assignment of the site with the minimum fitness and its first neighbors. In our simulations, the transient times of different lattice sizes for the system to achieve the statistically stationary state are attained by observing the slope value of the time evolution of the Hamming distance until it reaches to a fixed value as suggested in Ref. \cite{BB-EPJ}. The chosen transient times are $4\times10^{5}$, $1\times10^{6}$ and $2.4\times10^{6}$ for $N=50\times50$, $N=100\times100$ and $N=200\times200$, respectively.
\begin{figure}[t] 
   \centering
   \includegraphics[height=.35\textheight]{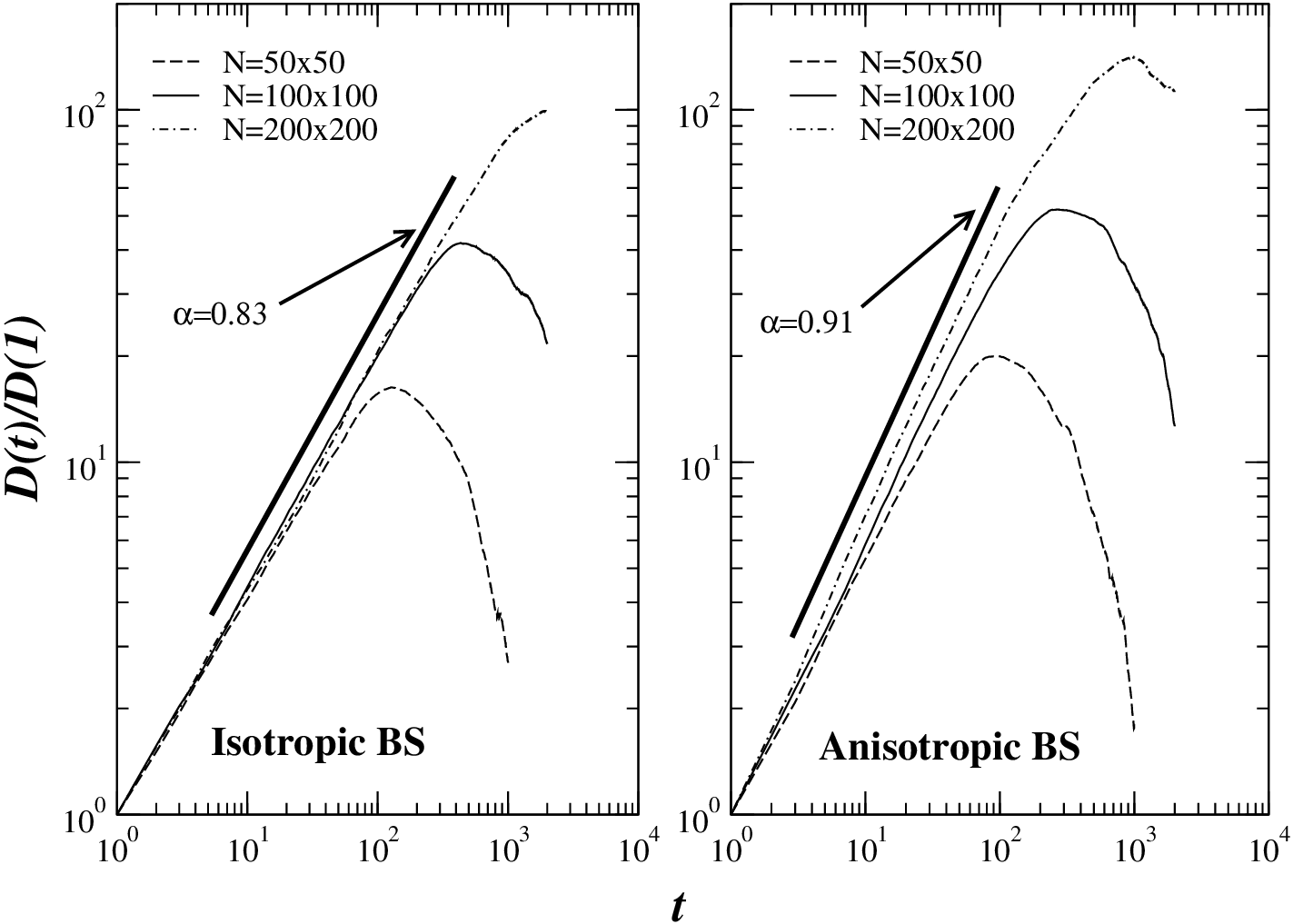} 
   \caption{The short-time evolution of the normalized Hamming distance between an equilibrium configuration and its slightly modified replica for three different system sizes of the isotropic (left) and anisotropic (right) BS models. After the initial power-law growth $D(t)\propto t^{\alpha}$ where $\alpha=0.83\pm0.03$ for the isotropic case and $\alpha=0.91\pm0.03$ for the anisotropic case \cite{BB-EPJ}, it decreases as both configurations converges to the same sequence of random numbers. The number of realizations used in calculations is $200$ for $N=50\times50$ and $N=100\times100$, and $100$ for $N=200\times200$.}
   \label{fig:short}
\end{figure}
     
Once the configuration achieves the statistically stationary state we implement the standard damage spreading algorithm given in Section~\ref{sec:Int}. After a characteristic time both configurations $f_{i,\,j}^{1}$ and $f_{i,\,j}^{2}$  will be composed by the same sequence of random numbers just shifted by a random distance. That is, $f_{i,\,j}^{1}$ and $f_{i,\,j}^{2}$ are now indistinguishable and can be considered as identical replicas. 

For such an implementation of the Bak-Sneppen model on a square lattice the Hamming distance measure $D(t)$ defined at each time $t$ as the smallest among the $L\times L$ possible values of $D_{k,\,\ell}(t)$ ($k=0,\,1,\,\cdots,\,L-1$ and $\ell=0,\,1,\,\cdots,\,L-1$) given by 
\begin{equation}
\label{eq:Hamming-UT2}
D_{k,\,\ell}(t)=\left\langle\frac{1}{N}\sum_{i,\,j=1}^{L}| f_{i,\,j}^{1}-f_{i+k,\,j+\ell}^{2}|\right\rangle.
\end{equation}
Notice that $D_{0,0}(t)$ corresponds to the measure used in previous damage spreading studies of $2$-dimensional BS model with the same scaling exponents \cite{BB-EPJ}. 

In Fig.~(\ref{fig:short}) our results for the short-time evolution of the normalized Hamming distance $D(t)/D(1)$ for three different system sizes $N=50\times50$, $N=100\times100$ and $N=200\times200$ are shown in the case of $2$-dimensional isotropic and anisotropic BS models. The initial power-law regime where $D(t)\propto t^{\alpha}$ extends for longer periods as the configuration size increases. The corresponding scaling exponents are $\alpha\simeq0.83$ for the isotropic case and $\alpha\simeq0.91$ for the anisotropic case and these $\alpha$ values are in good agreement with the previously reported values \cite{BB-EPJ}. After the initial power-law regime ends up at a characteristic time which depends on the system size, Hamming distance reaches a maximum and starts to decrease. The topological behavior of the short-time evolution of the Hamming distance in $2$-dimensional BS model is similar to the behavior obtained in $1$-dimensional BS model \cite{UT-IJMPC14}.
     
\begin{figure}[t] 
   \centering
   \includegraphics[height=.35\textheight]{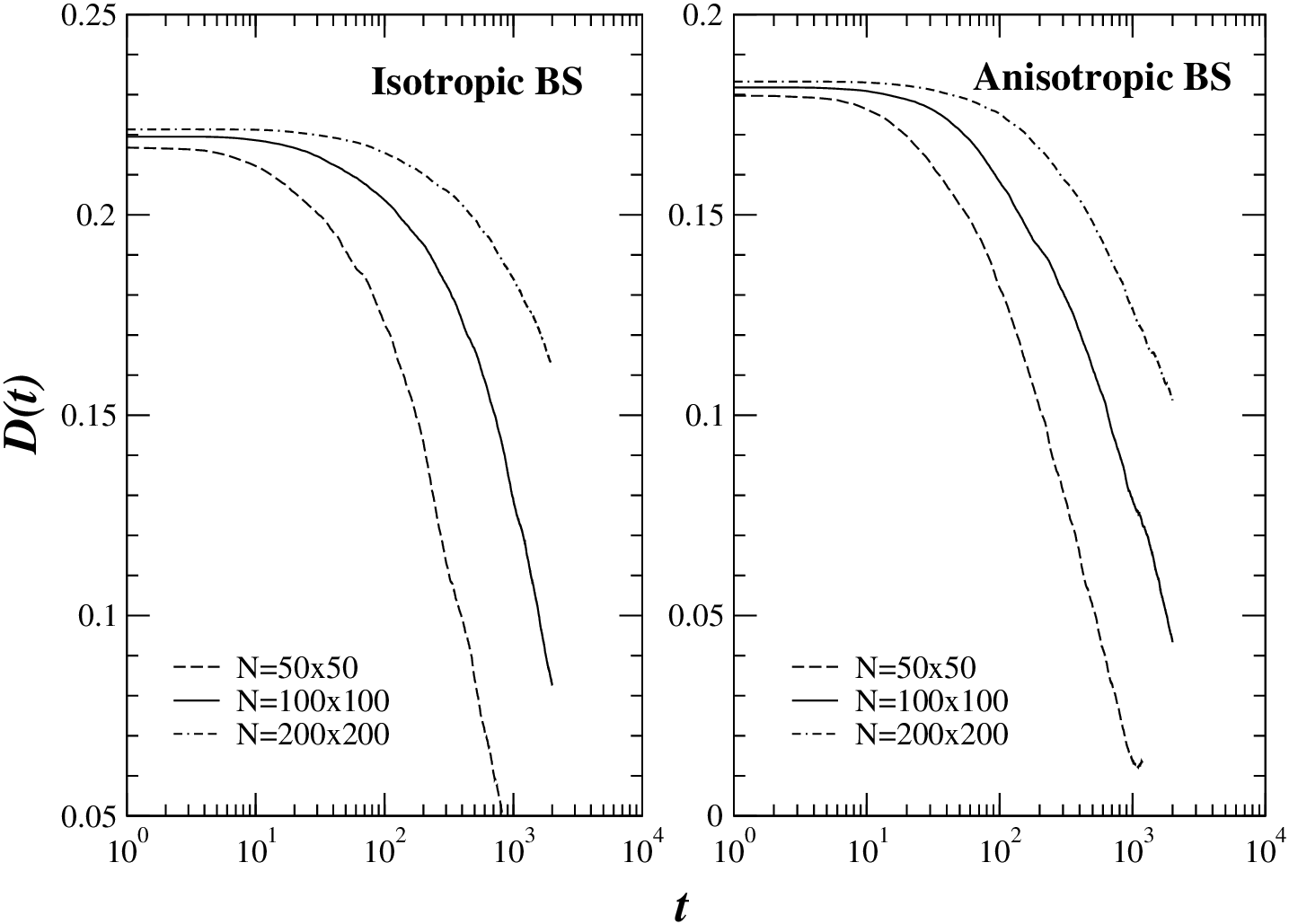} 
   \caption{The long-time evolution of the Hamming distance between two uncorrelated equilibrium configurations for three different system sizes. After the initial transient is achieved $D(t)$ exhibits a non-trivial decay followed by an exponential relaxation due to finite size effect. Additionally, from the figures one can notice that after the initial transient the Hamming distance is close to its initial value, $D(0)=0.224$ and $D(0)=0.187$ for the isotropic (left) and anisotropic (right) cases, respectively. The number of realizations used in calculations is $200$ for $N=50\times50$ and $N=100\times100$, and $100$ for $N=200\times200$.}
   \label{fig:long}
\end{figure}
In order to study the long-time dynamical regime of the Hamming distance evolution one needs to use a slightly different approach than the one mentioned above for the short-time dynamical regime. The new approach can be described as follows. Consider two initially uncorrelated random configurations of the BS model on a square lattice evolving under distinct noises. These configurations eventually reach uncorrelated statistically stationary states. For very large chains, it is well known that the Hamming distance between such configurations is $1/3$ of the width of the fitness distribution in equilibrium \cite{Valleriani-JPA32}. This fact remains true for the configurations described on a square lattice. Then the initial Hamming distance between $2$-dimensional two configurations of the isotropic and anisotropic BS models is given by
\begin{equation}
\label{eq:Hamming-D0}
D(0)=\sum_{i,\,j=1}^{L}| f_{i,\,j}^{1}-f_{i,\,j}^{2}|=\frac{1-f_{c}}{3}=
\left\{\begin{array}{l}0.224\qquad\text{(isotropic BS)}\\ \\0.187\qquad \text{(anisotropic BS)}.\end{array}\right.
\end{equation}
Once two uncorrelated configurations at their statistically stationary states are obtained we let both configurations evolve in time under the influence of the same external noise. That is, the minimum fitness and its nearest neighbors of each configuration are replaced by the same fitness values chosen from the same uniform random distribution in the interval $[0,\,1]$. As time increases, both configurations converge to the same sequence of fitness values while displaced in space. To investigate the convergence of both configurations, the Hamming distance measure defined as the minimum among all displaced distances $D_{k,\,\ell}(t)$ (see Eq.~(\ref{eq:Hamming-UT2})) is used. As emphasized before, making use of this Hamming distance measure enables one to observe the long-time evolution of the minimal Hamming distance. Such an observation is shown in Fig.~(\ref{fig:long}). As it is the case for $1$-dimensional case, the evolution of the minimum Hamming distance follows the solution of a generalized nonlinear equation, firstly proposed in the context of non-extensive thermostatistics \cite{TsallisJSP,TsallisLecture,GellMann-CT}, which includes in a single expression the transient, pre-asymptotic and asymptotic exponential relaxation regimes \cite{UT-IJMPC14,UT-PhysicaD}. From Fig.~(\ref{fig:long}) it is seen that the pre-asymptotic regime is very close to a slow logarithmic decay which is associated with the uncorrelated nature of the local variables in the statistically stationary states. Comparing our results of  $2$-dimensional BS model with $1$-dimensional case \cite{UT-IJMPC14,UT-Physica375} reveals that the long-time evolution of the minimum Hamming distance measure of both versions of the BS model exhibit topologically similar behavior.  
\section{\label{sec:Activity}Activity rate and waiting time distribution}
The smallest Hamming distance among all possible displacements determined by the fitness sites $\{k,\ell\}$ given in Eq.~(\ref{eq:Hamming-UT2}) is computed by comparing the state of two replicas introduced in the previous section. The optimal displacement determined by the fitness sites $\{k^{*},\ell^{*}\}$ evolve in time intermittently jumping between a few positions which provide almost the same Hamming distance. As it happens in $1$-dimensional BS model \cite{UT-Physica375}, after a long time evolution the optimal displacement between two $2$-dimensional replicas stops jumping (i.e., activity) once the replicas becomes identical without the need of any additional displacement. 
\begin{figure}[t]
   \centering
   \includegraphics[height=.35\textheight]{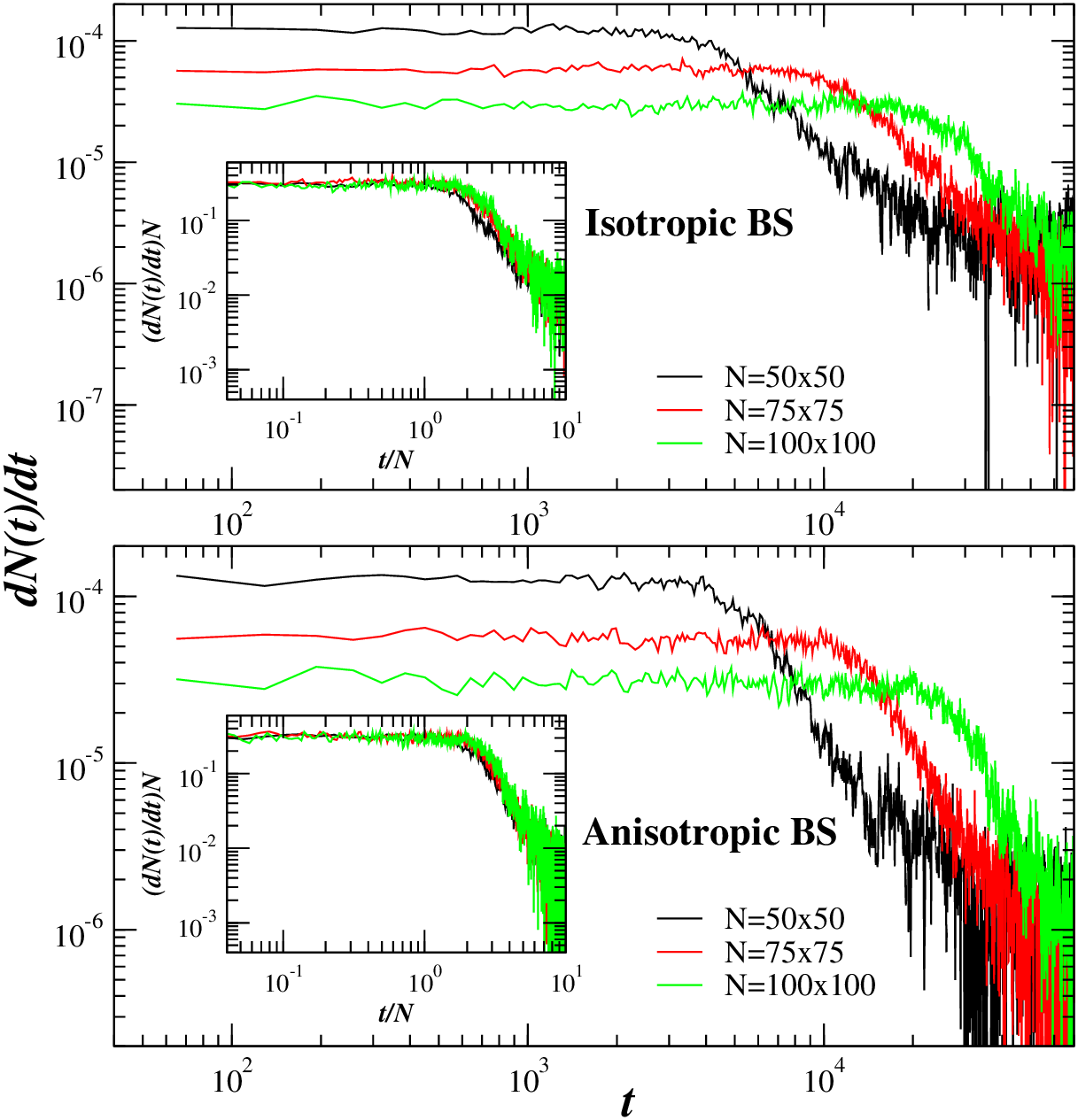} 
   \caption{(color online) The time evolution of the activity rate $dN/dt$ for three different system sizes of the isotropic (up) and anisotropic (down) BS models. During the initial regime, which depends on the system size the averaged activity rate is roughly constant crossing over to a regime of continuously decreasing activity. The initial regime scales as $1/N$ while the regime with the decreasing activity scales as $1/t$ (see insets). The number of realizations used in calculations is $200$ for $N=50\times50$ and $N=75\times75$, and $150$ for $N=100\times100$.}
   \label{fig:activity}
\end{figure}

To characterize this process, we have computed the activity rate $dN(t)/dt$, where $N(t)$ is the total number of jumps the optimal displacement has made until time $t$ for $2$-dimensional isotropic and anisotropic BS models. The number of realizations used in calculations is $200$ for $N=50\times50$ and $N=75\times75$, and $150$ for $N=100\times100$. At this point it should be emphasized that the computational difficulties (i.e., elapsed time to get the results) led us to work  with $100\times100$ lattice as the largest lattice size, because the time period needed to get the most realistic results for the activity rate and waiting time distributions increases rapidly as the lattice size increases. 

The physical interpretation of the picture given on Fig.~(\ref{fig:activity}) is as follows: the first regime where the time evolution of the activity rate is roughly constant giving a plateau corresponds to the initial relaxation towards the most predominant displacements providing the set of smaller Hamming distances. The activity rate scales as $1/N$, namely,
\begin{equation}\label{eq:scaling}
\frac{dN(t)}{dt}=\frac{1}{N}f\left(\frac{t}{N}\right). 
\end{equation}
During this period the two replicas stay uncorrelated and the fitness updates occur at uncorrelated random sites. This means that one can expect jumps to take place at a constant rate. These jumps match the minimal Hamming distance condition and evolve just a few displacements for which the Hamming distance is close to the smallest one. The second regime where the jump activity rate decreases is a consequence of the decrease in the minimal Hamming distance and reflects the locking of the two replicas at a predominant displacement. Configurational fluctuations of the time required to achieve the final displacement gives rise to the slowly decaying average activity rate characterized by a power-law on the form $dN(t)/dt\propto t^{-1}$. The size and time scaling of the jump activity rate is demonstrated in the insets of Fig.~(\ref{fig:activity}). 

Comparing our results in $2$-dimensional BS model with the results previously obtained in $1$-dimensional BS model \cite{UT-Physica375} reveals that the topology of the activity rate evolution in time is almost the same for both dimensions while the finite-size scalings are different. In our case the activity rate scales as $1/N$ while it scales as $1/N^{2}$ in $1$-dimensional case \cite{UT-Physica375}.            

\begin{figure}[t]
   \centering
   \includegraphics[height=.35\textheight]{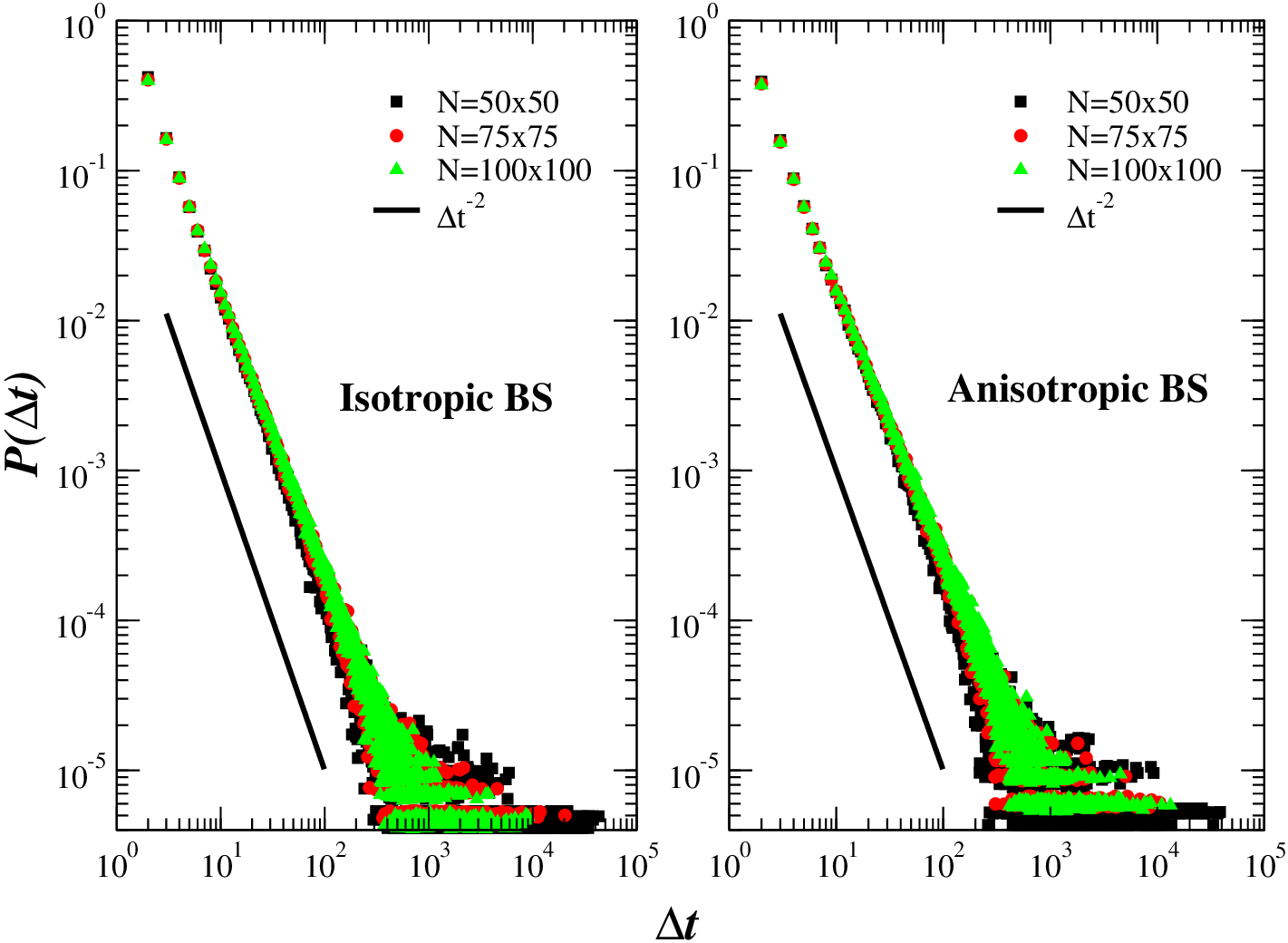} 
   \caption{(color online) The time evolution of the waiting times distribution for three different system sizes of the isotropic (left) and anisotropic (right) BS models. For both versions of the BS model it scales as $P(\Delta t)\propto\Delta t^{-2}$. The number of realizations used in calculations is $200$ for $N=50\times50$ and $N=75\times75$, and $150$ for $N=100\times100$.}
   \label{fig:pwt}
\end{figure}

Additionally, the waiting time distribution $P(\Delta t)$, i.e., the distribution of time intervals between two successive jumps is calculated to investigate the intermittence of the displacement jumps. Our results for $P(\Delta t)$ is shown in Fig.~(\ref{fig:pwt}). As it is the case for $1$-dimensional BS model, $P(\Delta t)$ follows a well defined power-law scaling with $P(\Delta t)\propto\Delta t^{-2}$ for both, isotropic and anisotropic cases. This means that the jumping process has no characteristic time scaling besides that due to finite-size effect.        

\section{\label{sec:Conc}Conclusion}
We have studied the convergence dynamics of two uncorrelated random configurations of the isotropic and anisotropic BS models on a square lattice by implementing the damage spreading technique. The chosen measurement way of the Hamming distance, which is defined as the smallest among $N$ possible values of $D_{j}(t)$ at each time step $t$ enabled us to capture both, the short-time and long-time evolution dynamics of the Hamming distance. 

It has been shown that the short-time evolution of the Hamming distance exhibits an initial power-law regime scaling as $D(t)\propto t^{\alpha}$, where $\alpha\simeq0.83$ for the isotropic and $\alpha\simeq0.91$ for the anisotropic cases. These $\alpha$ values are in full agreement with those previously obtained by using the conventional Hamming distance measure. After a characteristic time period, which depends on the system size, the initial power-law growth of the Hamming distance reaches a maximum and then follows a decreasing regime. In order to study the long-time dynamics of the Hamming distance we have considered two initially uncorrelated random configurations of the BS model on a square lattice. Our extensive simulation results have shown that the initial Hamming distance $D(0)$ attains the value of the $1/3$ of the width of the fitness distribution as it is the case in $1$-dimensional case, i.e., $D(0)=0.224$ for the isotropic and $D(0)=0.187$ for the anisotropic BS models characterized on a square lattice.

Considering the algorithm based on the definition of the minimal Hamming distance between all possible relative displacements of two uncorrelated configurations we have reported that the optimal displacement evolves in time intermittently. It has been seen that in the short time period where the two replicas stay uncorrelated the activity rate scales as $1/N$ following a decreasing regime in the long time period where the activity rate has a power law, $dN(t)/dt\propto t^{-1}$.   Moreover, the waiting time distribution was computed and shown to follow a power-law decay $P(\Delta t)\propto\Delta t^{-2}$. The scaling laws obtained for the time evolution of the activity rate and the waiting time distribution reflect the absence of typical spatio-temporal scales in the mechanism of displacement jumps used to quantify the convergence dynamics. 

The time evolution of the activity rate and the waiting time distribution for the isotropic and anisotropic versions of $2$-dimensional BS models reveal that they both evolve in time almost the same suggesting that choosing a preferred direction to propagate the avalanches does not have considerable influence on the convergence dynamics of two uncorrelated configurations of $2$-dimensional BS model. 

A difference between the BS models defined on a chain and $2$-dimensional lattice appears on the scaling factor defined in Eq.~(\ref{eq:scaling}). As it is seen, this factor is $1/N$ for $2$-dimensional BS model while it is $1/N^{2}$ in $1$-dimensional case. Although, the obtained scaling factors in $1$- and $2$-dimensional cases suggest a relation with the number of fitness in the lattice, a generalization of this relation to the higher dimensional cases needs further investigation.         
             
\acknowledgements
This work has been supported by TUBITAK (Turkish Agency) under the Research Project number 104T148 and by Ege University Research Fund under the research project 2006/BIL/026.
\bibliography{CD2dBS}
\end{document}